# Data objects and documenting scientific processes: An analysis of data events in biodiversity data papers

**(This paper has been accepted by the Journal of the Association for Information Science and Technology (JASIST))**

**Kai Li (College of Computing and Informatics, Drexel University, Philadelphia, PA; kl696@drexel.edu)**

**Jane Greenberg (College of Computing and Informatics, Drexel University, Philadelphia, PA; jg3243@drexel.edu)**

**Jillian Dunic (Department of Biological Sciences, Simon Fraser University, Burnaby, B.C.; jdunic@sfu.ca)**

**Abstract:** The data paper, an emerging scholarly genre, describes research datasets and is intended to bridge the gap between the publication of *research data* and *scientific articles*. Research examining how data papers report data events, such as data transactions and manipulations, is limited. The research reported on in this paper addresses this limitation and investigated how data events are inscribed in data papers. A content analysis was conducted examining the full texts of 82 data papers, drawn from the curated list of data papers connected to the Global Biodiversity Information Facility (GBIF). Data events recorded for each paper were organized into a set of 17 categories. Many of these categories are described together in the same sentence, which indicates the messiness of data events in the laboratory space. The findings challenge the degrees to which data papers are a distinct genre compared to research papers and they describe data-centric research processes in a through way. This paper also discusses how our results could inform a better data publication ecosystem in the future.

"As a cultural form, database represents the world as a list of items and it refuses to order the list. In contrast, a narrative creates a cause-and-effect trajectory of seemingly unordered items (events). Therefore, database and narrative are natural enemies. Competing for the same territory of human culture, each claims an exclusive right to make meaning out of the world." (Manovich, 1999)

# 1 Introduction

Data is an important element in the modern scientific enterprise, directly contributing to its basis of empiricism (Edwards, 2010; Gitelman, 2013). The historical relationship between science and data is being radically redefined by the new scientific paradigm in which data plays a central role in the scientific research (Hey, Tansley, Tolle, & others, 2009). The changing roles of data in science further invites us to relocate data in the conceptual and methodological frameworks adopted by scientists, in order to demonstrate its claimed values.

As one of the frames (i.e., metaphor) to understand data, *data publication* serves as a powerful vehicle for introducing data into existing scientific frameworks, as an answer to the aforementioned need. This metaphor transforms data into discrete, well-documented, publishable, and citable objects by borrowing concepts from scholarly publications (Parsons & Fox, 2013). A telling example of this metaphor is the emerging scientific genre of data papers. A data paper is a "scholarly publication of a searchable metadata document describing a particular online accessible dataset, or a group of datasets, published in accordance to the standard academic practices" (Chavan & Penev, 2011, p. 3). By directly using published papers as the proxy of data objects (Candela, Castelli, Manghi, & Tani, 2015), the quality of the metadata of datasets can be evaluated by other scientists before being published, a direct response to the claim that scientific data should be peer reviewed (Costello, Michener, Gahegan, Zhang, & Bourne, 2013; Mayernik, Callaghan, Leigh, Tedds, & Worley, 2015). This benefit partly contributes to the popularity of data papers among researchers in increasingly more scientific fields (Candela et al., 2015; Gorgolewski, Margulies, & Milham, 2013).

In terms of content, data papers focus on "information on the what, where, why, how and who of the data" (Callaghan et al., 2012, p. 112), rather than original research results (Kratz & Strasser, 2015). This distinct focus of data papers makes some researchers believe that this genre is an opportunity to bridge the gap between scientific texts and data reuse, as it supplies credible records of method metadata (Chao, 2015; Kratz & Strasser, 2014). Method-related information has proven to be an important source of trustworthiness of scientific data (Cragin, Palmer, Carlson, & Witt, 2010; Demchenko, Grosso, De Laat, & Membrey, 2013; Faniel & Jacobsen, 2010; Faniel & Zimmerman, 2011; Michener, Brunt, Helly, Kirchner, & Stafford, 1997; Zimmerman, 2007, 2008) and could greatly affect the ability of a data object to "travel" beyond its original context, an important standard to evaluate its success (Edwards, Mayernik, Batcheller, Bowker, & Borgman, 2011; Leonelli, 2016).

This positivistic view that data papers function as objective and detailed laboratory records concerning data objects is in contrast with theories on knowledge production developed from the field of science and technology studies (STS). Many researchers from this tradition have proven that a research paper is essentially a rhetorical device crafted to persuade other researchers to follow a proposed research agenda. Compared to this literary function, research papers may pay less attention to describing the scientific actions leading to the results in a comprehensive or even accurate manner (Knorr-Cetina, 1983; Latour & Woolgar, 1979; Swales, 1990).

In light of this dispute between positivism and constructivism, data papers are holding a unique position in our current scholarly communication system, because 1) data papers have distinct features from research articles but are nevertheless created under similar conditions; and 2) there are deep conflicts between data and narratives that data papers are claimed to bridge, as stated in the epigraph[1]. However, given the lack of empirical evidence, we are far from being able to properly evaluate data papers' positions in the contemporary scholarly communication. To offer a first piece of evidence towards filling this gap, research presented in this paper was undertaken to investigate both *how data events are described in the genre of data papers* and *what kinds of data events are described*. For the purposes of this work, data events are operationalized as any action during the research process that is directly and purposefully imposed on data and has resulted any change to the data object. This is a first step in understanding the relationship between data objects and scientific activities in the era of data-driven science.

---

[1] It should be noted that the gaps between narratives and data have been emphasized to be filled in the scopes of some data journals, such as the *Biodiversity Data Journal* (Smith et al., 2013).

The remainder of this article is organized as follows: The literature review provides theoretical background of the new data-driven scientific paradigm and the concept of data publication. This is followed by the presentation of our research questions and a detailed discussion of the research method and data. The key findings are reported, succeeded by a discussion. Finally, the conclusion specifically addresses how this study could inform future research and empirical works related to the publication of scientific data.

## 2 Literature review

### 2.1 Data-driven science, a new mode of scientific investigation

With the increasing amount of and dependency on data in scientific studies, a new mode of scholarship emerged by the end of the 20th century. This new scientific mode was famously named the "Fourth Paradigm" (Hey et al., 2009), as a new chapter of scientific research characterized by an overwhelmingly large amount of data. In this paradigm, research data becomes a first-class, if not the most important, research object. Moreover, the concept of "digital scholarship" (Unsworth, 2006) illustrates how this new mode of knowledge production has extended the scope of scientific research and challenged traditional scientific methodologies and epistemologies.

The expansion of scientific research is especially exemplified in the emerging field of data science, which takes scientific data as the direct object of investigation. In his highly-cited article published on Forbes.com, Press (Press, 2013) has stated that, despite earlier efforts to establish a scientific field focusing on data, the idea of *data science* has nevertheless become more established since the beginning of the 21st century, after the founding of *Data Science Journal* and the *Journal of Data Science* in 2002 and 2003, respectively. It should be noted that both journals share strong interests in the application aspects of data systems and works. This characteristic of data science is mirrored in Waller & Fawcett's famous definition of data science as an "application of quantitative and qualitative methods to solve relevant problems and predict outcomes" (Waller & Fawcett, 2013, p. 78). Moreover, there is a growing number of studies that focus on the practical aspects of scientific data, many of which fall into the concepts of *data curation* or *digital curation* (Higgins, 2011; McLure, Level, Cranston, Oehlerts, & Culbertson, 2014; Weber, Palmer, & Chao, 2012; Yakel, 2007).

Data science has been naturally introduced into many traditionally data-driven scientific fields, such as genomics and astronomy (Stephens et al., 2015). This trend is also present in the new coinages combining a scientific field name and -*informatics*, most notably bioinformatics (Ouzounis & Valencia, 2003; Trifonov, 2000) and cheminformatics (Noordik, 2004). Being strongly dependent upon large amount of data, biology has become an important site in which data and data science play significant roles in knowledge production. This, not surprisingly, coincides with biologists' embracing the practice of data papers, the topic of the present study.

## 2.2 Data papers as a form of data publication

Data publication is arguably the most important concept in the field of data science, as the availability of data is the prerequisite for any action that can be conducted on it (Costello et al., 2013, p. 6). Data publication, based on the influential discussion by Parsons & Fox (Parsons & Fox, 2013), is an metaphor based on the concepts from printed publication, so that research data can be properly understood, managed, and eventually, reused. It is the most mature metaphorical rendering of research data, even though it is still shrouded by many controversies (Borgman, 2015; Kratz & Strasser, 2014). One consensus around this topic is that data publication should entail mechanisms to make the data permanent, peer-reviewed, and citable, besides just making data available (Brase, 2009; Lawrence, Jones, Matthews, Pepler, & Callaghan, 2011; Wilkinson et al., 2016). This definition makes data publication strongly related but not necessarily distinguishable from *data sharing* on the one side and *data curation* on the other side. For example, Costello (2009) defined a hierarchy between data publication and data sharing in terms of the formality of sharing, even though both terms have been used interchangeably in some other cases (Enke et al., 2012). Moreover, Gray and colleagues (Gray, Szalay, Thakar, Stoughton, & vandenBerg, 2002) argued that data publication is data curation; while data curation is also perceived as a component of *data Publication* by Lawrence and colleagues (Lawrence et al., 2011).

Regardless of its definition, data publication is by no means a unitary system. Lawrence and colleagues (2011) identified five different approaches to *data Publication* by 1) how roles involved in data publication are distributed among actors; and 2) the relationships among data objects, data documentations, and publications citing data. For example, in the simplest model identified in the paper, *Stand Alone Data Publication*, new publications cite archived data objects per se. While in a number of other cases, such as *Publication by Proxy*, *Journal Driven*

*Data Archival*, and *Overlay Data Publication*, a *defining journal article* is published to describe the data and serve as the citable object; these three models are distinguishable by their different relationships between data and defining article. It can be argued that most, if not all, data papers are most similar to the model of *Publication by Proxy*, even though data papers have a stronger focus of not including any data analysis and results (Kratz & Strasser, 2014).

Data publication is a device to bridge the gap between human and computers within the lifecycle of research data (Lawrence et al., 2011). Metadata plays central roles towards achieving this goal, despite the facts that creating metadata is highly labor-intensive (Borgman, 2012; Peer & Green, 2012) and its creation is deeply situated in individual communities with own needs (Edwards et al., 2011; Mayernik, Batcheller, & Borgman, 2011). In an effort to classify metadata for data archives, Lawrence and colleagues (Lawrence, Lowry, Miller, Snaith, & Woolf, 2009) categorized four groups of metadata serving different purposes: browse metadata, archive metadata, character metadata, and discovery metadata. A chief functional requirement for the description of data beyond all these categories is the methods to produce the described data objects. Method related information has been proven to be an essential factor for researchers to trust and reuse data objects created by others (Cragin et al., 2010; Demchenko et al., 2013; Faniel & Jacobsen, 2010; Faniel & Zimmerman, 2011; Wallis, Rolando, & Borgman, 2013). It is important to point out here that, in fact, detailed methodological descriptions were not the most direct motivation for data papers. However, method was identified as a core element in the original design of this academic genre (Chavan & Penev, 2011), and data papers were subsequently perceived to be a chance to bridge the gap between natural language and structured data for method information (Chao, 2015; Chavan, Penev, & Hobern, 2013; Gorgolewski et al., 2013).

As part of this trend, several gaps need to be addressed before the role of data papers are fully endorsed in our scholarly communication system. One particular issue is that data papers adopts the traditional scholarly publication pipeline. This approach inevitably traps data papers into the classic debates concerning the socio-narrative functions played by scholarly publication, although it reduces the cost of reinventing a new wheel. On the one hand, arguments have been put forth to posit data papers in the same family of research articles. These arguments include that data papers are created under similar social conditions with research papers. For example, data papers bear the expectation to be published (and thus awarded) under the scrutiny of reviewers, which is a major source of the constructivist arguments about research papers (Latour, 1987; Latour & Woolgar, 1979). Moreover, data papers are also subject to space limitations and the community's norms, which makes it impossible for authors to include every technical detail (Bazerman, 1988; Berkenkotter & Huckin, 2016; Knorr-Cetina, 1983). Despite the aforementioned arguments, as a metadata document, data papers are supposed to offer enough information, so that data objects can overcome the friction to be reused (Chao, 2015; Edwards et al., 2011). Moreover, data journals normally require authors to include comprehensive technical details. For example, *Scientific Data* ask researchers to "describe any steps or procedures used in producing the data, including full descriptions of the experimental design, data acquisition assays, and any computational processing (e.g. normalization, image feature extraction)."[2] A similar example is *Zookeys*, which states that authors should "provide sufficient information to allow someone to repeat your work."[3]

Given the clear conflicts between both conditions of data papers and our lack of empirical knowledge about data papers, it is not only important, but urgent, to investigate what information is described in data papers regarding the methods to create and manipulate the data objects. This is the major source of motivation for the present study.

# 3 Research questions

To advance our understanding of data papers as an increasingly important venue in scholarly communication, we pursued the following two research questions concerning what and how data-related workflows and processes are described in this genre in the present study:

**Research question 1: What data events are described in data papers?**

This question focuses on identifying the functional categories of data events covered by data papers. A data event is defined as any action during the research process that is directly and purposefully imposed on data and has resulted any change to the data object per the description of data papers.

**Research question 2: How are data events covered in data papers?**

By relying upon the classification of data events, this question aims to understand how each category of data events is covered in data papers. Moreover, we also examined the cases in which data events of different types are described in the same sentence, as a reflection of the close relationship (or "package") between these events.

---

[2] https://www.nature.com/sdata/publish/submission-guidelines#methods
[3] https://zookeys.pensoft.net/about#Authors-Guidelines

# 4 Method and Procedures

In order to pursue the research questions discussed above, a content analysis was conducted to identify data events from full texts of data papers. Identifying themes is an essential function of content analysis (Berg, 2004; Strauss & Corbin, 1990); in the case of the present analysis, we are trying to identify and classify data events as the theme from data papers as the artefact. Following this step, data events were classified based on their functions using an inductive approach.

## 4.1 Data sample

We examined the full collection of 82 data papers from a list of data papers both curated by and about the Global Biodiversity Information Facility (GBIF). GBIF is a major biodiversity research data repository supporting creation of data papers based on the metadata of deposited datasets (Moritz et al., 2011) and was considered a mature source for our study. All the papers in the list were retrieved from corresponding database on July 1, 2017. This list has recently been removed from the GBIF website after an upgrade of the website; however, it is still available through the WayBack Machine service of Internet Archive[4]. The paper list and its bibliographic data have been deposited on Figshare (Li, 2018). In the present study, all these data papers will be referred to by their paper ID listed in dataset.

All data papers in our sample were published between 2011 and 2017. Most of these papers (72 out of 82 papers) are from six journals (Figure 1); each of the other 10 papers is from a separate journal. It should be noted that *Zookeys*, *Biodiversity Data Journal*, and *PhytoKeys* are also the three journals published by Pensoft Publishers, who collaborated with GBIF to evaluate the feasibility of data paper publication (Chavan & Penev, 2011).

### Figure 1: Frequencies of top journals in the sample

---

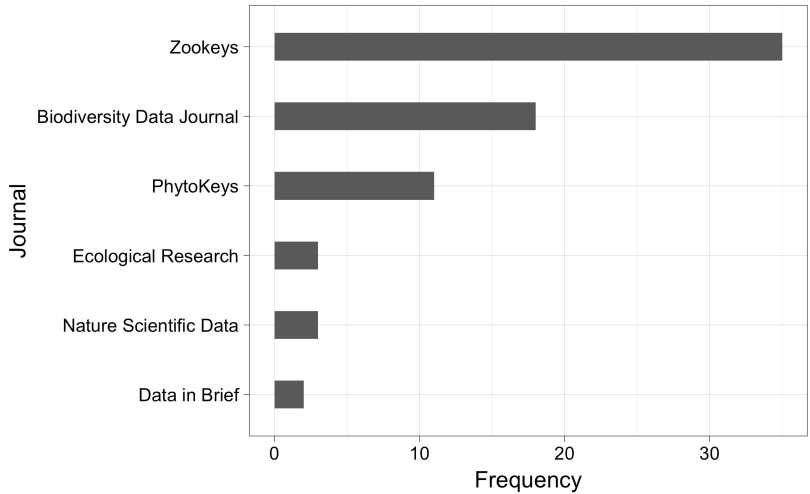

## 4.2 Procedures to identify and classify data events

The content analysis involved two key steps, guided by a set of principles. The steps outlined below were conducted by examining the full text of the 82 papers, that comprised the study's sample. The outline of the procedures also includes scenarios that help to illustrate and document the research steps:

**Step 1: Data event identification.** We first identified all sentences in the sampled papers with any description of data events based on their working definition specified in Research Question 1. We only considered unstructured texts in the article body, rather than structured descriptive information. An example of the structured description in a data paper is quoted below. Often presented in tables, such information can be seen as a visual representation of metadata of the data object, which functions differently from narrative texts.

"Design Type(s): data integration objective • observation design •biodiversity assessment objective." (Paper 1)

**Step 2: Data event classification.** Based on the identified sentences, we classified all data events using an inductive approach. By using the open coding method, two coders from different academic backgrounds (namely, information science and ecology) separately identified sentences with data events and developed the classification scheme based on the identified sentences. The classification schemes and the corresponding categorization were reviewed and integrated by both coders and an extra coder from information science. Differences were resolved based on discussions among all coders. During the classification processes, the following principles were followed as much as possible.

**Principle 1: The classification is based on the functional view of data events.** Data events should be classified based on what goals an action is intended to achieve. This approach is consistent with the aim of this research to understand which different data events are described in data papers. Moreover, this view also makes it inevitable that the taxonomy of the classification shares many similar terms with existing data lifecycle models (Ball, 2012).

**Principle 2: The classification is NOT based on presumptions about the sequence between data events.** No time sequence should be presumed between the actions to be classified, unless a sequence is explicitly described in the papers. The purpose of this principle is to have an empirical recording of the data objects based on what is described in scientific texts, which distinguishes the present work from data lifecycle models.

Our approach to data events makes it possible that events belonging to different categories could be packaged in the same sentence with or without their sequence explicitly expressed. An example of each situation is shown below:

Scenario 1: "In these cases, the geospatial information has been checked and herbarium specimens have been reviewed to confirm taxonomic identification." (Paper 2)

Scenario 2: "After identification, specimens were randomly selected to be re-examined by either first or last author in order to check identification accuracy." (Paper 15)

In the first example, the events of validating geospatial information ("the geospatial information has been checked") and reviewing taxonomic identification ("herbarium specimens have been reviewed to confirm taxonomic identification") can be assumed to happen at the same time, given that no other information is available in the text. In the second case, the review of identification ("specimens were randomly selected to be re-examined") happened after the identification was made. However, in both cases, we determined that the co-appearance of data events in the same sentence is an indication of their conceptual closeness with each other. Based on this assumption, we specifically examined all such sentences in this research.

## 5 Results

The results are reported in the following two parts: the first part concerns the categories of data events; the second is an analysis of the co-occurrence of different data event categories.

## 5.1 Data event classification

In total, 767 sentences containing any data event were identified in our sample. These sentences scatter across 79 papers in our sample; we could not find any such sentence in three papers. Among these sentences, 932 data event instances were identified. Based on this sample, our final data event classification scheme is presented in Table 1, with an example for each category.

**Table 1: Categories of data events**

| Category | Abbreviations | Definition | Example |
|---|---|---|---|
| Data analysis | Dana | The activities in which statistical and/or analytical procedures are applied to collected data to acquire new information. | "The cardinality, a combination of abundance and vegetation cover, of each species within each plot was estimated using the Wilmanns scale (Reichelt and Wilmanns 1973)." (Paper 33) |
| Data classification | DCla | The activities of assigning categorical values to the data based on criteria other than taxonomical ones. | "Nine types of habitat and landscape were defined to collect accurate information on the habitat preference of all ant species (Dekoninck et al. 2005)." (Paper 65) |
| Data collection | DCol | The activities in which data are collected from the natural world or other data sources into the dataset described in the paper. | "All (and only) indications given on the label have been recorded." (Paper 39) |
| Data formatting | DFor | The design and/or change of the data structure based on metadata schemas. | "We used Darwin Core Archive Validator tool (http://tools.gbif.org/dwca-validator/) to check whether the dataset met Darwin Core specifications." (Paper 43) |
| Data identification | DIde | The assigning of identifiers to observations in the dataset. | "Unique collections' accession numbers were assigned to each specimen." (Paper 54) |
| Data modification | DMod | The change of the scale used in the original data object or adding new fields describing data management-related steps. | "In those specimens with more precise coordinates in the label, the coordinates have been generalized to blur sensitive information due to the threatened degree of these taxa." (Paper 2) |

| Data registration | DReg | The creation of a list of observations in a local or shared environment. | "We first built a preliminary list of locality descriptions by searching for a list of keywords (e.g. 'Botanical garden')." (Paper 9) |
|---|---|---|---|
| Data removal | DRem | The deletion of observations from the dataset. | "10 % of the data were removed through this process due to inaccurate GPS coordinates." (Paper 24) |
| Data sharing | DSha | The sharing of the data object in any public venue. | "The dataset with metadata was uploaded to the Integrated Publishing Toolkit (IPT) of the Colombia node of Global Biodiversity Information Facility (GBIF)." (Paper 48) |
| Data validation | DVal | The checking and correcting of data points for quality control purposes. | "The accuracy and geographic coordinates of localities where collection specimens came from was checked based on geological grounds (BRGM sources)." (Paper 63) |
| Data visualization | DVis | The creation of visual representations based on the data values. | "Distribution maps have been created using ArcGis 10.2." (Paper 2) |
| Databasing | Dat | The creation or changing of the physical architecture of the database. | "Data were then exported and ingested into KeEMU by the NHM database team." (Paper 17) |
| Georeferencing | Geo | The association of spatial locations with a map or geological names. | "All SIFlore records are georeferenced through the code of the corresponding grid square." (Paper 35) |
| Metadata creation | Meta | The creation of descriptive metadata for the dataset. | "The metadata from the dataset have been completed directly in the IPT." (Paper 46) |
| Taxonomy identification | TaxIde | The activities in which taxonomic categories and/or scientific names are assigned to specimens. | "Taxonomic identification was performed at the Italian Antarctic National Museum (MNA, Section of Genoa) and at the British Antarctic Survey (BAS) laboratories." (Paper 70) |

| Taxonomy validation | TaxVal | The checking and correcting of taxonomic identification or scientific names based on taxonomic standards. | "Accurate spelling of scientific names and taxonomic synonyms was revised according to Arnedo (2003)." (Paper 12) |
|---|---|---|---|
| Tool developing | Tool | The development of tools to support the lifecycle of the described data object(s). | "A customized application (app) (Figure 4) was built to facilitate both data collection and storage (Pérez-Pérez et al. 2013 - http://obsnev.es/noticia.html?id=4513)." (Paper 43) |

Figure 2 shows the total frequency of each data event category on the levels of sentence (the top bar) and paper (the bottom bar). *Data collection* is the largest category on both levels: 217 instances were identified in 71 papers. Its large size can be attributed to two factors. First, the method of collection is arguably the most basic information about any data object, which should be recorded in any metadata document. Second, we purposely chose not to further separate this category into smaller units, such as digitalization and data recording, in order to have a comprehensive category corresponding to the data lifecycle model as a basis for a direct comparison between these two approaches to recording data-related information. Notwithstanding its significance for the data object and any future use of it, the collection of data fails to be mentioned in eight of the 79 papers. Likewise, other categories are even less represented in the data papers we examined. This result is a direct challenge to the notion that data papers are objective and impeccable vehicles of the "actual" life history of data objects.

**Figure 2: Frequency of each data event category**

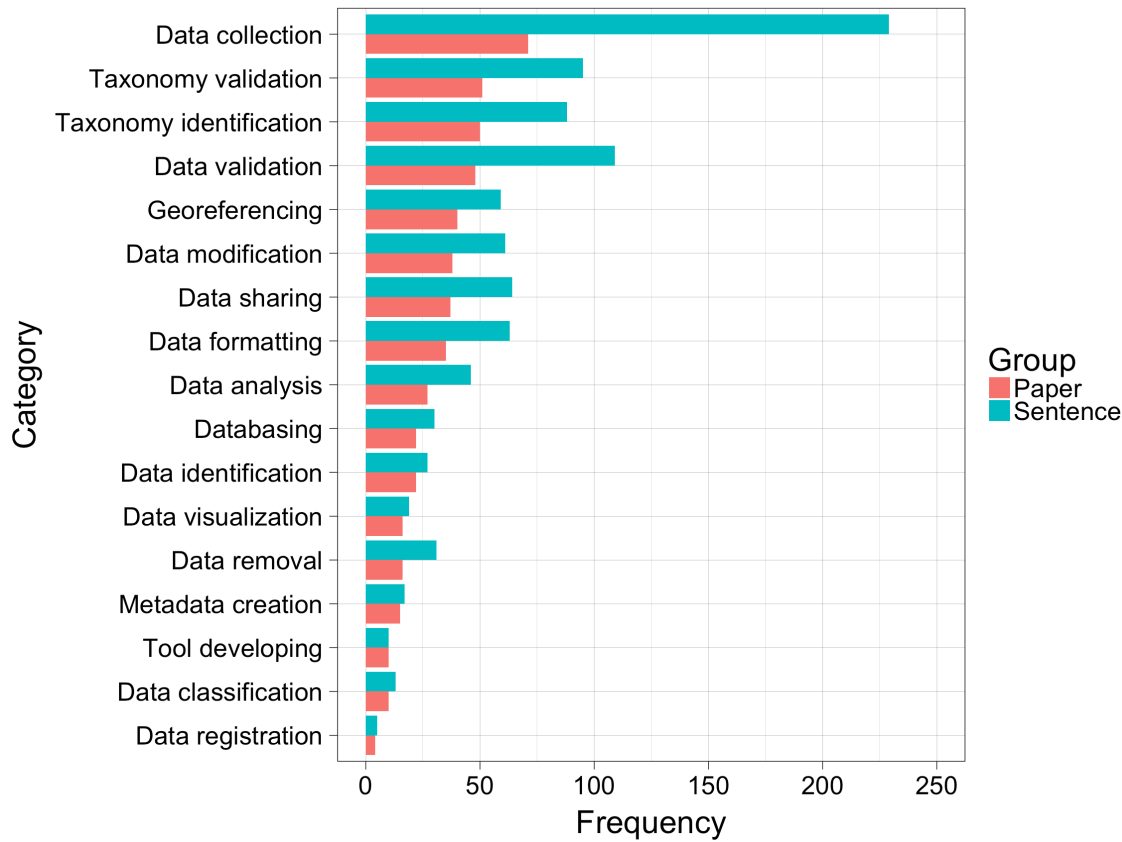

Another notable category is *Data analysis*. In many cases, this category clearly points to functions to be fulfilled by research papers, such as:

"Bootstrap analysis was also carried out to know the evolutionary history of bacteria." (Paper 22)

The existence of this information in data papers raises questions about the definitional distinctions between research papers and data papers, and how such differences are perceived and operated on by researchers in reality. These questions necessitate future comparative studies evaluating how scientific processes are represented differently in these two academic genres.

## 5.2 Co-occurrence of data event categories

In order to understand the relationship between individual data events, we analyzed the sentences in which data events belonging to different categories are mentioned together and recorded what data event categories are involved into such data event pairs. 309 data events from 144 sentences were identified. Figure 3 shows the frequency of each category pair. This graph demonstrates the existence of a wide array of data events that can be packaged in the same sentence. This is especially true for *Data collection*, given its large number of instances; but it is also the case for smaller categories such as *Tool developing* and *Data registration*. Four key category pairs whose frequency is at least 10 are discussed below, in order to illustrate the meanings of such connections.

**Figure 3: Sentence-level co-appearance of data events belonging to different categories (Abbreviations offered in Table 1)**

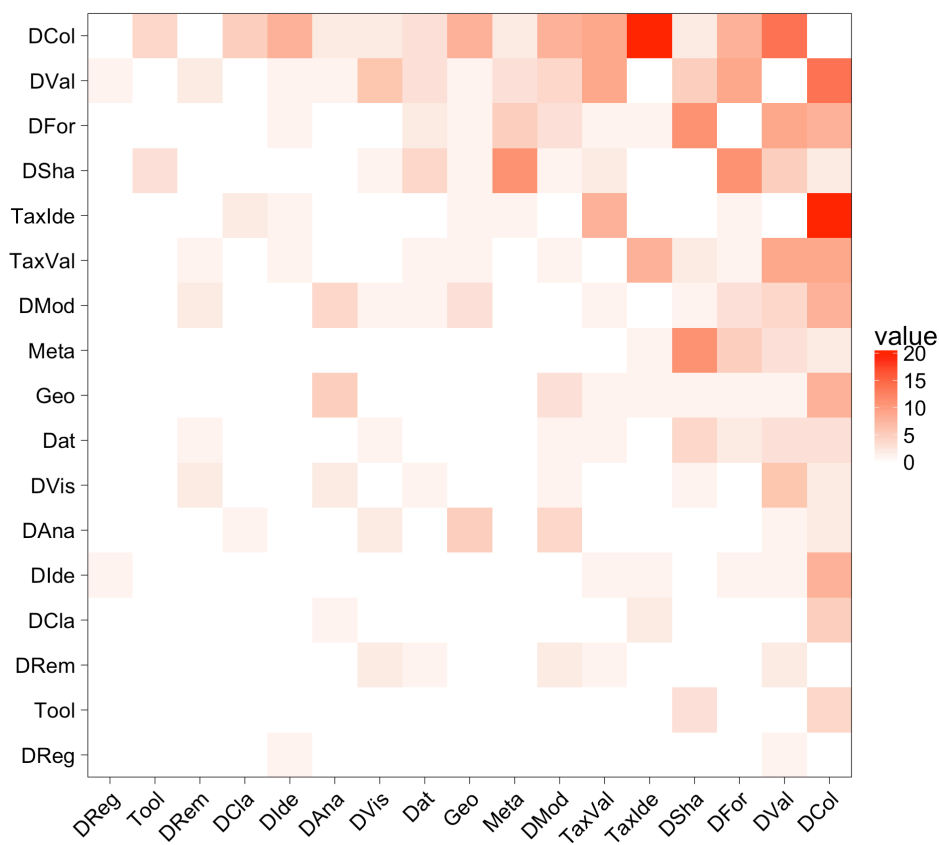

**Data collection & Taxonomy identification:** Data events belonging to these two categories are frequently packaged together as a general description of how data were gradually acquired from the specimens before it is digitized into a dataset. Most of these descriptions are staged in the physical laboratories. For example (highlight ours):

"The captured individuals were anaesthetized with tricaine methanesulfonate (MS-222; Sigma Chemical Co., St. Louis, MO) or 2-phenoxyethanol, **identified**, **counted and measured** (total length in millimetres, and, in some cases, weight in grams)." (Paper 23)

**Data collection & Data validation:** Based on the sampled data papers, validation procedures were commonly mentioned as occurring right after data were collected. For example (highlight ours):

> "Then, the resulting DarwinCore records were completed by adding the ImageURL and TypeStatus fields, after which **it was validated with the DARWIN TEST tool** (Ortega-Maqueda and Pando 2008)." (Paper 46)

**Data sharing & Data formatting:** *Data formatting* and *Data sharing* are naturally two steps found close to each other in the texts, since both steps are closely connected to the final production of a dataset. For example (highlight ours):

> "The dataset is **exported to DarwinCore v1.2 format** and **uploaded to the IPT of the GBIF Spanish node** (http://www.gbif.es:8080/ipt)." (Paper 74)

**Data sharing & Metadata creation:** Metadata creation is also more likely to be related to the later stages of dataset development, rather than its "raw" form. However, as shown in the example below (highlight ours), the connections between these two steps are further strengthened by the function of the GBIF IPT, software used to publish datasets on GBIF.

> "The Integrated Publishing Toolkit (IPT v2.0.5) (Robertson et al. 2014) of the Spanish node of the Global Biodiversity Information Facility (GBIF) (http://www.gbif.es/ipt) was used both to **upload the Darwin Core Archive** and **to fill out the metadata**." (Paper 44)

Besides these examples, it should be noted that many, if not more, instances of the relationship between data events are expressed by the logical connections between sentences and cannot be observed by our method. Despite the fact that this is beyond the scope of the present study, it is definitely worth investigating in future studies.

## 6 Discussion

As a foundation for this research, a classification scheme for the function of data events was developed in this study. We believe this classification scheme, being developed from an empirical and functional perspective, is an important contribution of the present study. Seventeen types of data events were identified from the 82 data papers that were examined. Many data events belonging to different categories were described by researchers in the same sentence, which indicates that these categories are perceived to be close to each other by researchers. This observation confirms that the overall data processes are messier and more contingent than those depicted in data lifecycle models, and individual data events can be connected to each other in various unexpected ways. Given the limitations of data lifecycle models based on our results and their importance in various data and research management applications (Demchenko et al., 2013; Frey, 2008), it is worth considering the possibilities of augmenting existing data lifecycle models with empirically-driven evidence in the future.

In addition to this implication, the classification activity for this research helped us to identify challenges to assumptions about the roles played by data papers in scholarly communication under the new data-driven mode of scientific knowledge production. The two most obvious challenges target the incomplete coverage of the data processes in data papers and the possibly blurry boundaries between data papers and research papers. These challenges are elaborated as follows.

**Data papers fail to completely cover data events.** This challenge is evident in our finding that events related to the collection of the data object, a task that covers various actions and nearly indispensable for any dataset, fail to occur in eight papers, let along other less-mentioned activities. This conclusion should not be construed as a negation of the importance of data papers as a link between data and text, the two most important types of media in digital scholarship. However, it does question the degree to which one should take data papers as a reliable source of method-related information about data objects (Chao, 2015; Kratz & Strasser, 2014).

**There are functional overlaps between data papers and research papers.** The second challenge to data papers arises in our finding that a large number of data events in data papers, especially those in the category of *Data analysis*, bear functions that are supposed to be achieved by research papers. This mixed mode of data publication has long existed and was categorized as *Publication by Proxy* by Lawrence and colleagues (Lawrence et al., 2011) before the advent of the genre of data papers, in which case data objects are reported along with other findings in the research paper. Building upon the history of data publication, our findings suggest that there are still functional overlaps between data papers and traditional research papers, most fundamentally derived from the metaphorical foundation of data papers. This functional overlap between these genres will be an interesting topic for future studies on the rhetorical functions of data papers.

Both challenges point to the fact that new developments are inevitably built on existing infrastructures and practices. This point, of course, echoes Borgman's argument that treating data as publication could cost us the opportunities to explore new models of scholarly communication (Borgman, 2015). Despite its advantages, such as higher user familiarity and greater easiness to attract rewards, data papers, based on our analysis, are far from a perfect mechanism to communicate method and procedures information related to data objects. A number of researchers have proposed that sharing data as software can offer better version control for data objects and make data more accessible (Parsons & Fox, 2013; Schopf, 2012). These considerations lead to the following suggestions about the future roles data papers could play in the scholarly communication landscape.

It should be expected that the future of scholarly communication will remain diverse, if not more so than today (Stuart, 2017): more than one approaches to sharing and publishing data and its workflows will co-exist across different scientific fields, or even within the same research community. As demonstrated by its popularity (Candela et al., 2015), data papers have become an established approach adopted by scientists. This makes it important to think about how to improve this genre and/or reexamine its roles, rather than totally dismissing its values.

As a genre built upon natural languages, data papers are primarily a human-readable document, much less designed for reproducing data workflows in computational approaches. Comparing with digital scientific workflow objects (Taylor, Deelman, Gannon, & Shields, 2014), this genre both shares some essential functions and has significantly different focuses. This relationship between data papers and scientific workflow objects enables the possibilities in which these objects play complimentary, rather than competitive, roles in the future. Given their different nature, they inscribe different aspects of the data workflow towards different types of users (i.e., human readers for data papers and computers for workflow objects) and they are published in different venues. However, their creations and presence can be further coordinated and linked together so that the data object being described by both objects can be better understood and reused.

One example of the coordination is to augment data papers with structured information concerning scientific workflows. This structure could be a journal-recommended checklist, covering the most basic steps in common data lifecycles, so that writers can be given the chance to reflect on what data events they have decided to include in the article. But a more promising solution to bridge data papers and other formal representations of the data workflow is to use ontologies or taxonomies to augment data papers. An especially relevant resource is the PROV Data Model (PROV DM) recommended by W3C (Moreau & Missier, 2013). PROV DM is a generic data model describing provenance information, i.e., "record that describes the people, institutions, entities, and activities involved in producing, influencing, or delivering a piece of data or a thing" (Moreau & Missier, 2013, p. 3). Specifically, it defines relationships among entities, activities, and agents, which are at the core of using and producing entities, and offers mappings to other metadata schemas, such as XML and Dublin Core. Besides PROV DM, a number of ontologies or ontological-driven systems concerning research or data workflows have been developed during the past decade, most of which are based on the use case of managing and using a raw dataset in a digital system (Haston, Cubey, Pullan, Atkins, & Harris, 2012; Korolev & Joshi, 2014; McPhillips, Bowers, Zinn, & Ludäscher, 2009; Stephan et al., 2010; Wombacher, 2010). As shown in our analysis, data events described in data papers also cover those steps in which data is collected from specimen. Despite limitations of these models, they could serve as good starting points for data papers to be annotated and understood by machine agents.

From a different direction, we also believe one of the future uses of data papers is a data source from which structured information concerning data events are extracted or learnt, at least before sharing research or data lifecycle objects becomes an established norm. Building ontologies from natural language corpora has become a popular paradigm during the past 20 years (Liu, Hogan, & Crowley, 2011; Shamsfard & Barforoush, 2004; Uschold & Uschold, 1996). This approach has all the potentials to transform our knowledge about the roles played by data and data publications in the production and communication of scientific knowledge. As the first step of this process, more applied linguistics studies need to be conducted on the genre of data papers, so that we can gain enough ground truth about the interactions of language, research objects, and scientific knowledge in these new publications.

## 7 Conclusions and next steps

The research reported here offers an initial analysis of what data-related methods are described in the emerging genre of data papers. Our results reveal that information concerning method recorded in data papers should not be taken for granted as entirely accurate descriptions of the actual laboratory process. This issue needs to be regarded as a central one in the scholarly communication of data-driven science. This finding is consistent with reflections on the limitations of the metaphor of data publication (Borgman, 2015; Parsons & Fox, 2013). Most importantly, the research presented here and in other noted research may indicate that treating data as print publication could cost us opportunities to develop a scholarly communication system more suitable for the nature of data (Borgman, 2015, p. 278). These considerations point the importance of sharing more structured representation of scientific workflow (Corcho et al., 2012; Sompel, Payette, Erickson, Lagoze, & Warner, 2004). However, as an increasingly popular practice, we should not negate the values of data papers, but to relocate it as a complementary source of method-related information concerning data, that is mainly towards human readers and NLP techniques to develop new ontologies about data workflows.

Despite the possibilities and evidence offered in this paper, we fully acknowledge the idiosyncrasies of our classification due to the similar topics and authors of our sample. To address this challenge, we used the method of open coding (Berg, 2004) and different knowledge backgrounds of coders, to extract data events and their categorizations based on the scientific text along, rather than preexisting assumptions about what data is supposed to go through in scientific research. This approach to data events is consistent with our goal to understand the *in-situ* processes of data objects as seen in data papers.

Based on the results and limitations presented above, much work is needed to fully understand the practice and impact of data papers. Three directions are the most interesting related to the topic of the present paper. The first direction is a comparative analysis of data papers across disciplinary boundaries. The composition of data papers is inevitably situated in the specific epistemic communities (Cronin, 2003; Knorr-Cetina, 1999). Thus, it would be an important research agenda to find deeper connections among research methods, scientific language, and the scientific discipline. The second direction concerns what roles tools have played in the creation of data objects, based on the descriptions of data papers. The third direction is to better understand data papers as a socio-narrative device, by using applied linguistic methods: this is also an important preparation before we can extract new knowledge concerning data workflows from the natural language in data papers. We are hoping the present paper will be the first step towards a comprehensive introduction of data into the scholarly communication system.

## Acknowledgement

The project is part of the RDA/US Data Share fellowship sponsored through a grant from the Alfred P. Sloan Foundation #G-2014-13746.